\documentstyle[B_Phys,12pt]{article}
\def\thspace{\kern.08333em}
\def\beq{\begin{equation}}
\def\eeq{\end{equation}}
\def\sss{\scriptscriptstyle}

\def\bd{B_d^0}
\def\bdb{\overline{B_d^0}}
\def\bs{B_s^0}
\def\bsb{\overline{B_s^0}}

\def\lft{{\sss L}}
\def\rht{{\sss R}}
\def\leapprox{\lower.6ex
              \hbox{\kern+.2em$\buildrel <\over\sim$\kern+.2em}}
\def\bsg{b\to s\gamma}
\def\bsll{b\to s\,\ell^+\ell^-}
\def\xs{x_s}
\def\fbb{f^2_{B_d}B_{B_d}}
\def\fbbs{f^2_{B_s}B_{B_s}}
\def\fbs{f_{B_s}}

\def\npb#1#2#3{{\em Nucl.~Phys.} {\bf B#1} (19#2) #3}
\def\plb#1#2#3{{\em Phys.~Lett.} {\bf #1B} (19#2) #3}

\def\prd#1#2#3{{\em Phys.~Rev.} {\bf D#1} (19#2) #3}

\def\prl#1#2#3{{\em Phys.~Rev.~Lett.} {\bf #1} (19#2) #3}

\def\zpc#1#2#3{{\em Zeit.~Phys.} {\bf C#1} (19#2) #3}

\def\semilepmodels{{1}}
\def\HQET{{2}}
\def\hadmodels{{3}}
\def\gronwak{{4}}
\def\otherproc{{5}}
\def\langacker{{6}}
\def\houwyler{{7}}
\def\Afb{{8}}
\def\LEPrightB{{9}}
\def\bsgQCD{{10}}
\def\ali{{11}}
\def\CLEO{{12}}
\def\Hewett{{13}}
\def\Barger{{14}}
\def\BelangerGeng{{15}}
\def\SUSYcorrs{{16}}
\def\SUSYbsg{{17}}
\def\ARGUSCLEO{{18}}
\def\bsllnunurate{{19}}
\def\bsllrate{{20}}
\def\aligreub{{21}}
\def\UA1{{22}}
\def\AliMor{{23}}
\def\BBM{{24}}
\def\Fran{{25}}
\def\GrinNir{{26}}
\def\Buras{{27}}
\def\CampOdon{{28}}
\def\HiggsBmumu{{29}}
\def\Randall{{30}}
\def\penguins{{31}}
\def\Deshpande{{32}}
\def\CHill{{33}}
\def\CQuigg{{34}}
\def\Bcmass{{35}}
\def\Bcproduction{{36}}
\def\Bcdecay{{37}}
\def\BcdecayQCD{{38}}
\def\BcCP{{39}}
\def\GroWyler{{40}}
\def\xsreview{{41}}
\def\fBlattice{{42}}
\def\fBratio{{43}}
\def\fourthgen{{44}}
\def\SUSYxs{{45}}
\def\lrxs{{46}}
\def\zfcnc{{47}}
\def\BCPviolation{{48}}
\def\Valencia{{49}}
\def\BBVmodel{{50}}
\def\Btriple{{51}}

\begin{document}
\rightline{UdeM-LPN-TH-93-178}
\rightline{October, 1993}
\vskip1.5truecm
\centerline{\large\bf B DECAYS IN THE STANDARD MODEL}
\centerline{\large\bf AND BEYOND}
\vskip1em
\centerline{DAVID LONDON}
\centerline{\em Laboratoire de physique nucl\'eaire, Universit\'e de
Montr\'eal}
\centerline{\em C.P. 6128, succ. A, Montr\'eal, Qu\'ebec, CANADA, H3C 3J7.}
\vskip1.5truecm

%\maketitle
 \pagestyle{plain}    % with page numbers

\onehead{1.}{INTRODUCTION}

When Vera L\"uth asked me to give a talk on $B$ decays in and beyond the
standard model (SM), I readily accepted. However, when I sat down and made
a list of the topics I would have to cover, I quickly realized that I had
bitten off more than I could chew. My list consisted of the following
subjects:
\begin{itemize}
\item Semileptonic $B$ decays: these are typically described in one of two
ways. Either one picks a specific model,$^\semilepmodels$ or one uses the
Heavy Quark Effective Theory (HQET);$^\HQET$
\item Hadronic $B$ decays: such decays are usually described by the BSW
model;$^\hadmodels$
\item Right-handed $B$ decays:$^\gronwak$ the suggestion here is that $B$
decays are mediated not by the ordinary $W$, but rather by a right-handed
$W_\rht$;
\item Rare $B$ decays: included are the flavour-changing neutral-current
decays \\
$\bsg$, $\bsll$, $b\to s\nu{\overline\nu}$, $b\to sg$, $b\to sq{\bar q}$,
and $B^0\to \ell^+\ell^-$, as well as hadronic penguins ($B\to K\pi$,
etc.), and $B^0\to \gamma\gamma$;
\item the decay $B_u^+ \to\ell^+\nu$;
\item Exotic $B$ states such as $B_c$'s and $\Lambda_b$'s;
\item $B$-${\overline B}$ mixing -- $x_d$ and $x_s$;
\item T Violation (triple products),
\end{itemize}
and I'm sure I've overlooked some other possibilities. Given the length of
this list, I realized that I would have to limit myself to a subset of the
above topics. I therefore decided to discuss only right-handed $B$ decays,
certain rare $B$ decays, $B_c$ decays, $B_s^0$-${\overline{B_s^0}}$ mixing,
and T violation. Some of the other subjects, such as HQET and $B$ baryons,
are discussed elsewhere in these proceedings.$^\otherproc$ \newpage

\onehead{2.}{RIGHT-HANDED B DECAYS}

Gronau and Wakaizumi$^\gronwak$ (GW) have suggested that $B$ decays might
in fact be mediated by a right-handed $W_\rht$, instead of the SM
left-handed $W$. This possibility is predicated on two facts. First, the
chirality of $B$ decays has not yet been measured. And second, the mass of
the $W_\rht$ could still be relatively small:$^\langacker$
\beq
M_\rht^g \equiv \left({g_\lft \over g_\rht}\right) M_\rht > 300~{\rm GeV}~,
\eeq
where $M_\rht$ is the mass of the $W_\rht$, and $g_\lft$ and $g_\rht$ are
the left and right couplings, respectively.

With this in mind, GW have proposed a model in which the SM $W$ doesn't
couple to $B$'s at all. They interpret the long $B$ lifetime as being due
to the heaviness of the $W_\rht$, {\it not} to the smallness of $V_{cb}$.
That is,
\beq
\beta_g \equiv \left({g_\rht^2 \over g_\lft^2}\right)
\left({M_\lft^2 \over M_\rht^2}\right)
\sim \vert V_{cb} \vert = 0.044 \pm 0.006.
\eeq
Phenomenologically, $\beta_g$ is bounded to be $<0.07$.

In order for this model to be viable, the form of the right-handed CKM
matrix $V^\rht$ must take into account a large number of phenomenological
constraints involving $B$'s -- the $B$ lifetime, $b\to u$ transitions,
2-body $B$ decays, Cabibbo-suppressed $B$ decays, $\bd$-$\bdb$ mixing --
as well as the $K_{\sss L}$-$K_{\sss S}$ mass difference, $\Delta m_{\sss
K}$. The forms suggested by GW for both the left- and right-handed CKM
matrix, consistent with the above data, are
\beq
V^\lft = \left( \matrix{ \cos\theta_c & \sin\theta_c & 0 \cr
			-\sin\theta_c & \cos\theta_c & 0 \cr
			            0 &            0 & 1 \cr}\right)~,~~~~~
V^\rht = \left( \matrix{ c^2 & -cs & s \cr
	      s(1-c)/\sqrt{2} & (c+s^2)\sqrt{2} & c/\sqrt{2} \cr
	     -s(1-c)/\sqrt{2} & -(c-s^2)\sqrt{2} & c/\sqrt{2}\cr}\right),
\label{GWsolution}
\eeq
in which $\theta_c$ is the Cabibbo angle, and $s\equiv\sin\theta^\rht$,
$c\equiv\cos\theta^\rht$. The magnitude of $s$ is determined from $\vert
V_{ub}/V_{cb}\vert$, i.e.
\beq
s=0.08 \pm 0.02~.
\eeq
With this choice of left- and right-handed CKM matrices, all known data can
be explained with
\beq
M_\rht^g = {\hbox{300--600 GeV}}.
\label{wrightrange}
\eeq

In fact, strictly speaking, this is not completely true -- additional
assumptions are necessary. For example, this model demands the existence of
a right-handed neutrino with a mass $m(\nu_\rht) < m_b - m_c$. Furthermore,
if the $\nu_\rht$ is very light, muon decay experiments require either that
$M_\rht^g$ be in the upper part of the range of Eq.~\ref{wrightrange}, or
that the $\nu_\rht$ be unstable. In addition, from direct searches for
right-handed $W$'s at hadron colliders, the limit $M_\rht>520$ GeV is
obtained for $g_\rht=g_\lft$. Thus, if one wants a value for $M_\rht^g$ in
the lower part of the range of Eq.~\ref{wrightrange}, it is necessary that
$g_\rht$ be larger than $g_\lft$. Nevertheless, despite these caveats, the
model is interesting in the sense that it points out certain aspects of $B$
decays which must be examined in order to fully test the SM.

One possibly bothersome aspect of the GW solution (Eq.~\ref{GWsolution}),
pointed out by Hou and Wyler$^\houwyler$ (HW), is that $V_{cd}^\rht$ is
unnaturally small ($=0.0003$). One way to avoid this is to parametrize the
right-handed CKM matrix using two angles $\theta_{12}$ and $\theta_{13}$.
HW propose$^\houwyler$
\beq
V^\rht = \left( \matrix{ 1 & 0 & 0 \cr
	                 0 & 1/\sqrt{2} & 1/\sqrt{2} \cr
	                 0 & -1/\sqrt{2} & 1/\sqrt{2} \cr}\right)
\left( \matrix{ c_{13} & 0 & s_{13} \cr
	             0 & 1 & 0 \cr
	       -s_{13} & 0 & c_{13} \cr}\right)
\left( \matrix{ c_{12} & -s_{12} & 0 \cr
	        s_{12} & c_{12} & 0 \cr
	        0 & 0 & 1 \cr}\right),
\label{solutionI}
\eeq
and take $s_{12}\simeq 0.098$, $s_{13}\simeq 0.085$, in which case
$V^\rht_{cd} = s_{12} - c_{12} s_{13} \simeq 0.01$. I will refer to this as
solution (I).

HW also point out that even if the $b\to c$ transitions are dominated by
right-handed currents, $b\to u$ decays might still be mediated mainly via
left-handed currents. They thus arrive at solution (II):
\beq
V^\lft \simeq \left( \matrix{ 1 & \lambda & \delta \cr
		     -\lambda & 1 & \delta \cr
 	             - \delta & -\delta & 1 \cr}\right)~,~~~~~
V^\rht \simeq \left( \matrix{ 1 & -\varepsilon & \varepsilon \cr
	      \varepsilon & 1/\sqrt{2} & 1/\sqrt{2} \cr
	     -\varepsilon & -1/\sqrt{2} & 1/\sqrt{2}\cr}\right),
\label{solutionII}
\eeq
with $\lambda=\sin \theta_c$, $\delta\sim 0.05$ and $\varepsilon < 0.01$.

Now, the question is, how can one rule out these models? One of the
advantages of a hadron collider, as compared to an asymmetric $e^+e^-$
collider operating at the $\Upsilon(4s)$ resonance, is that one can search
directly for new physics. It is more likely that physics beyond the
standard model -- supersymmetry, extra Higgses, technicolour, etc. -- will
be first found via direct searches than by looking for indirect signals in
$B$ physics. As such, the most straightforward way to rule out models of
right-handed $B$ decays is simply to look for, and fail to find, a light
$W_\rht$.

Another possibility$^\houwyler$ is to look at certain $B$ decays which are
suppressed in these models relative to the SM. For example,
\begin{eqnarray}
{BR(b\to c{\bar c}d) \over BR(b\to c{\bar c}s)} & = & \lambda^2 \simeq
0.05, ~~~~~ {\rm (SM)}, \nonumber \\
& = & O(10^{-7}), ~~~~~~ {\rm (GW)}, \nonumber \\
& \leapprox & O(10^{-4}), ~~~~~~ {\rm (I,II)}.
\end{eqnarray}
In this case, if right-handed currents were responsible for $B$ decays, the
ratio of the decay rates of $B\to D^{(*)} D_s^{-(*)}$ and $B\to D^{(*)}
D^{-(*)}$ would differ from that of the SM. Similarly,
\begin{eqnarray}
{BR(b\to c{\bar u}s) \over BR(b\to c{\bar u}d)} & = & \lambda^2 \simeq
0.05, ~~~~~~~~~ {\rm (SM)}, \nonumber \\
& \simeq & 0.008, ~~~~~~~~~~~~~ {\rm (GW,I)}, \nonumber \\
& \simeq & \varepsilon^2 \leapprox O(10^{-4}), ~~~~~ {\rm (II)}.
\end{eqnarray}
Here one should compare, for example, ${\overline{B}}\to D^{(*)}\rho$ and
${\overline{B}}\to D^{(*)} K^*$.

Finally, there is the possibility of measuring the chirality of $B$ decays.
The lepton forward-backward decay asymmetry $A_{fb}$ in the decay
${\overline B}\to D^* \ell^- {\overline{\nu}}_\ell$ is sensitive to the
chirality of the $b\to c$ coupling.$^\Afb$ However, not being
parity-violating, $A_{fb}$ also depends on the chirality of the lepton
current, and therefore cannot distinguish models of right-handed $B$ decays
from the standard model. On the other hand, experiments at LEP {\it can}
make such a distinction. One looks$^\LEPrightB$ at the reaction $e^+e^- \to
Z^0 \to \Lambda_b X$, in which the $\Lambda_b$ is highly polarized, its
spin carried essentially entirely by the $b$-quark. The electron energy
spectrum in $\Lambda_b\to charm$ semileptonic decays is then quite
sensitive to the $V\pm A$ nature of the $b\to c$ coupling. In this way it
might be possible to rule out models of right-handed $B$ decays at LEP.

\newpage

\onehead{3.}{RARE $B$ DECAYS}

\ttwohead{3.1}{$\bsg$ (and $b\to d\gamma$)}

The flavour-changing decay $\bsg$ occurs first at one loop, and is
dominated at lowest order by the $t$-quark contribution:
\beq
{\cal M}(\bsg) = {G_{\sss F} \over \sqrt{2}} \, {e\over 4\pi^2} \,
\lambda_t F_2(x_t) q^\mu \epsilon^\nu {\bar s} \sigma_{\mu\nu}
\left( m_b (1+\gamma_5) + m_s (1-\gamma_5) \right) b~,
\label{bsgammarate}
\eeq
in which $\lambda_t \equiv V_{tb} V_{ts}^*$, $x_t=m_t^2/M_{\sss W}^2$, and
\beq
F_2(x) = {x \over 24 (x-1)^4}
\left[ 6x(3x-2) \log x - (x-1)(8x^2+5x-7)\right].
\eeq
However, this process receives important QCD contributions,$^\bsgQCD$ as
shown in Fig.~1. For example, for $m_t=150$ GeV, we find
\begin{eqnarray}
BR (\bsg) & = & 1.4 \times 10^{-4}~~~~~({\hbox{no QCD corrections}}),
\nonumber \\
& = & 4.2 \times 10^{-4}~~~~~({\hbox{including QCD corrections}}).
\end{eqnarray}

\vspace{6cm}
\centerline{Figure 1: Branching ratio for $\bsg$ in the SM with
(solid line) and without}
\centerline{\qquad\qquad\qquad~~~~ (dashed line) QCD corrections
(from Ref.~\ali\ (reproduced by permission)).}

\vskip0.2truecm
The rate for $b\to d\gamma$ is obtained from that for $\bsg$
(Eq.~\ref{bsgammarate}) by replacing the $s$-quark variables by $d$-quark
variables. Thus, to lowest order,
\beq
{BR (b\to d\gamma) \over BR (\bsg)} = \left\vert {V_{td}\over
V_{ts}} \right\vert^2~.
\eeq
However, there are additional corrections due to the breaking of
$SU(3)_{flavour}$. Estimating these, and taking into account the
uncertainty in the magnitude of $V_{td}$, one finds$^\ali$
\begin{eqnarray}
BR (\bsg) & = & {\hbox{3--5}} \times 10^{-4}~, \nonumber \\
BR (b\to d\gamma) & = & {\hbox{0.5--3}} \times 10^{-5}~.
\end{eqnarray}

Although the inclusive decay rate for $\bsg$ can be calculated with good
precision, it is well-known that exclusive decays are poorly understood
theoretically:
\beq
R_{\sss B} \equiv {BR (B\to K^*\gamma) \over BR (\bsg)} =
{\hbox{4--40 \%.}}
\eeq

CLEO has measured both inclusive and exclusive flavour-changing
decays:$^\CLEO$
\begin{eqnarray}
BR (\bsg) & < & 8.4 \times 10^{-4}~~~~(1991), \nonumber \\
		  & < & 5.4 \times 10^{-4}~~~~(1993), \nonumber \\
BR (B\to K^*\gamma) & = & (4.5 \pm 1.5 \pm 0.9) \times 10^{-5}~~~~(1993).
\label{CLEObounds}
\end{eqnarray}
These measurements have important consequences for models of new physics.

First consider models with two Higgs doublets (2HDM). In general, such
models will lead to flavour-changing neutral currents. This then requires
that the Higgs bosons be very heavy, rendering their effects in $B$ physics
unobservable. There are two ways to avoid this, distinguished by the
couplings of the fermions and the Higgses. One possibility (model I) is
that one Higgs doublet, $\phi_2$, gives mass to all fermions, while the
other doublet, $\phi_1$, decouples. In the other case (model II), one
doublet, $\phi_2$, couples to all $u$-type quarks, while the second Higgs
doublet, $\phi_1$, gives mass to $d$-type quarks. It is model II which
appears in supersymmetric and axion models.

In either of these 2HDM there are new contributions to the decay $b\to
s\gamma$, found by replacing the $W^\pm$ in the loop by a charged Higgs,
$H^\pm$. In these models, both Higgs doublets acquire vacuum expectation
values, denoted $v_1$ and $v_2$. We define $\tan\beta\equiv v_2/v_1$, which
is apriori completely free. The transition amplitude is then proportional
to
\beq
A_{\sss W} \left({m_t^2\over M_{\sss W}^2}\right) + \lambda A_{\sss H}^1
\left({m_t^2\over M_{{\sss H}^\pm}^2}\right) + {1\over \tan^2\beta}
A_{\sss H}^2 \left({m_t^2\over M_{{\sss H}^\pm}^2}\right)~,
\eeq
where $A_{\sss W}$ and $A_{\sss H}^{1,2}$ represent the SM and
charged-Higgs contributions to the amplitude, respectively. In model I,
$\lambda=-1/\tan^2\beta$, while $\lambda=+1$ in model II.

{}From this we see that in model I, there is an enhancement to the rate for
$\bsg$ only for small values of $\tan\beta$. In model II, the rate is also
enhanced for small $\tan\beta$. More importantly, due to the $A_{\sss H}^1$
term, the rate is {\it always} larger than that of the SM. This leads to a
lower bound on the mass of the charged Higgs in this
model,$^{\Hewett,\Barger}$ independent of the value of $m_t$. In Fig.~2,
taken from Ref.~\Hewett, the constraints on models I and II are shown for
$m_t=150$ GeV, using the 1991 CLEO bound (Eq.~\ref{CLEObounds}).

\vspace{6.1cm}
\centerline{Figure 2: Excluded regions in the
$M_{{\sss H}^\pm}$-$\tan\beta$ plane for models I and II, }
\centerline{\qquad\qquad~ for $m_t=150$ GeV, (from Ref.~\Hewett\
(reproduced by permission)).}

\vskip0.2truecm
For model I, we see that there is no $\tan\beta$-independent lower limit on
$M_{{\sss H}^\pm}$ coming from the bound on $\bsg$. However, in model II,
we find that $M_{{\sss H}^\pm} > 110$ GeV at large $\tan\beta$, with
stronger bounds for smaller values of $\tan\beta$. For model II this lower
limit has been updated$^\BelangerGeng$ using the 1993 data on $\bsg$
(Eq.~\ref{CLEObounds}): $M_{{\sss H}^\pm} >$ 320 GeV (540 GeV) for $m_t =$
120 GeV (150 GeV). This new lower bound has several important consequences.
First, the decay $t\to b H^+$ is no longer allowed. Second, if the two
Higgs doublets are part of a supersymmetric theory, the difficult region
for Higgs searches is now ruled out (see below, however). Finally, this
eliminates most large effects in 2HDM in other rare $B$ decays.

The implications of the limits on $BR(\bsg)$ are less clearcut for
supersymmetric models. If the main new contributions to $\bsg$ came from
the two Higgs doublets, then the constraints would be as described above.
However, the situation is more complicated. First, electroweak radiative
corrections to the charged-Higgs mass and to the charged
Higgs-fermion-fermion vertex can be substantial.$^\SUSYcorrs$ These
corrections tend to weaken the constraints on the charged-Higgs mass as a
function of $\tan\beta$. More importantly, in the minimal supersymmetric
standard model (MSSM), the contributions to $\bsg$ from other
supersymmetric particles may not be negligible.$^\SUSYbsg$ In this case
there can be cancellations with the charged-Higgs contributions, possibly
resulting in a branching ratio for $\bsg$ which is {\it smaller} than that
of the SM. Thus, it is impossible to say anything concrete regarding the
constraints on SUSY models due to $BR(\bsg)$.

Finally, left-right symmetric models are essentially unconstrained by the
limits on $BR(\bsg)$ (Eq.~\ref{CLEObounds}). Models with right-handed $B$
decays predict a rate for $\bsg$ which is down by a factor of 2 compared to
the SM. And in models with manifest left-right symmetry, the $W_\rht$ must
be so heavy that its effects in $\bsg$ are negligible.

\ttwohead{3.2}{$\bsll$}

In the SM, at the quark level, the decay $\bsll$ arises through penguin
diagrams with a virtual $\gamma$ or $Z^0$, as well as through box diagrams.
In addition, in contrast to $\bsg$, $\bsll$ receives important
long-distance contributions. These effects are dominated by the decays
$B\to \Psi (\Psi') X \to \ell^+\ell^- X$, whose branching ratios have been
measured by the ARGUS and CLEO collaborations$^\ARGUSCLEO$ to be
$O(10^{-3})$. The long-distance effects are then very important when the
$\ell^+\ell^-$ pair has an invariant mass close to that of the $\Psi$ or
$\Psi'$. However, since the long-distance contribution is so much larger
than the short-distance contribution, which is estimated to be $O(10^{-5})$
(see below), one has to worry about residual effects in the spectrum away
from the $\Psi$ and $\Psi'$ resonances. In other words, the invariant
dilepton mass spectrum is important in analysing $\bsll$.

The short-distance contributions have been
calculated:$^{\bsllnunurate,\bsllrate,\aligreub}$
\begin{eqnarray}
BR (B\to X_s \, e^+e^-) & = & {\hbox{0.6--2.5}} \times 10^{-5}~, \nonumber
\\
BR (B\to X_s \, \mu^+\mu^-) & = & {\hbox{3.5--14.0}} \times 10^{-6}~,
\end{eqnarray}
for 100 GeV $< m_t <$ 200 GeV. Note that the $m_t$-dependence is much more
important here than in $\bsg$. Also note the the UA1 upper limit:$^\UA1$
\beq
BR (B\to \mu^+\mu^-X) < 5 \times 10^{-5}~.
\label{UA1limit}
\eeq
The short-distance contributions for the inclusive decays $b\to
d\,\ell^+\ell^-$ have also been computed,$^\aligreub$ assuming $\vert
V_{td} / V_{ts} \vert = 0.21$:
\begin{eqnarray}
BR (B\to X_d \, e^+e^-) & = & {\hbox{2.6--10.0}} \times 10^{-7}~, \nonumber
\\
BR (B\to X_d \, \mu^+\mu^-) & = & {\hbox{1.5--6.0}} \times 10^{-7}~.
\end{eqnarray}

Again, it must be remembered that the above cross sections are only the
short-distance contributions. One can try to also include the long-distance
effects, but there are large uncertainties. Nevertheless it is possible to
isolate the short-distance contributions by looking at the forward-backward
asymmetry in the decay. In Fig.~3 one sees the angular distribution of the
decay, for three different values of $m_t$, in which $\theta$ is defined as
the angle between the momentum of the $B$-meson and that of the $\ell^+$ in
the centre-of-mass frame of the dilepton pair, and ${\hat s}$ is the scaled
dilepton invariant mass. This figure is taken from Ref.~\AliMor, to which I
refer the reader for more details.

\vspace{7.4cm}
\centerline{Figure 3: The angular distribution $d^2\, BR/dz\, d{\hat s}$ in
the decay $\bsll$, }
\centerline{~\, for ${\hat s}=0.3$ (from Ref.~\AliMor\ (reproduced by
permission)).}

As mentioned earlier, the constraints from $\bsg$ on two-Higgs-doublet
models preclude large enhancements to $\bsll$. As to supersymmetric models,
in Ref.~\BBM, it is found that the rate for $\bsll$ can be greater than
that of the SM by up to a factor of 2, when the electroweak symmetry is
broken radiatively. On the other hand, this reference predates the recent
CLEO bounds on $\bsg$, and I'm not sure how their inclusion would change
the predictions of SUSY models for $\bsll$. The feeling seems to be that
the CLEO data probably now precludes SUSY enhancements to $\bsll$, but this
should be checked.$^\Fran$

Another type of new physics which could lead to an enhancement of the rate
for $\bsll$ is extended technicolour. In fact, for certain models,
specifically those which include a ``techni-GIM'' mechanism, the
enhancement is too large.$^\GrinNir$ In such models, barring delicate
fine-tuned cancellations, the prediction for $BR(B\to \mu^+\mu^-X)$ is
$O(10^{-4})$, which is in conflict with the UA1 bound (Eq.~\ref{UA1limit}).
These models therefore appear to be ruled out. On the other hand, extended
technicolour models without a GIM mechanism are still allowed -- they
predict $BR(B\to \mu^+\mu^-X) = {\hbox{1--3}} \times 10^{-5}$, an
enhancement of roughly a factor of 4 compared to the SM.

\ttwohead{3.3}{$b\to s\nu{\overline\nu}$}

Although the decay $b\to s\nu{\overline\nu}$ has negligible QCD
corrections, it is very sensitive to the value of $m_t$. In the SM, its
branching ratio is calculated to be$^{\bsllnunurate,\aligreub,\Buras}$
\beq
\sum_i BR (b\to s {\bar\nu}_i \nu_i) = {\hbox{2.8--13.0}} \times 10^{-5}
\eeq
for 100 GeV $< m_t <$ 200 GeV.

This branching ratio is not expected to be significantly affected by the
presence of new physics. In two-Higgs-doublet models, any possible effects
are already ruled out by the $\bsg$ measurement, and the inclusion of
supersymmetric particles$^\BBM$ is not expected to lead to any enhancement.

\ttwohead{3.4}{$B_s^0\to \mu^+\mu^-/\tau^+\tau^-$}

In order to deduce the form of the operator leading to the decay $B_s^0 \to
\ell^+\ell^-$, one notes the following points. First, the $s$-$b$ matrix
element is
\beq
\langle 0 \vert {\bar s} \gamma^\mu \gamma_5 b \vert B \rangle = f_{\sss B}
P_{\sss B}^\mu~.
\eeq
This is because the matrix element of ${\bar s} \gamma^\mu b$ vanishes due
to considerations of parity (the $B_s^0$ is a pseudoscalar) and ${\bar s}
\sigma^{\mu\nu} b$ won't work since there aren't enough Lorentz vectors to
construct a scalar. Second, $P_{\sss B}^\mu \, {\bar u}_\ell \gamma_\mu
v_\ell = 0$, which means we need a helicity flip in the leptonic current.
Thus, the operator describing the decay $B_s^0 \to \ell^+\ell^-$ is
\beq
{\cal O} \sim {\bar s}\gamma^\mu \gamma_5 b \, {\bar\ell} \gamma_\mu
\gamma_5 \ell~.
\eeq
The helicity flip means, of course, that the final answer will depend on
the lepton mass.

The branching ratio for the decay $B_s^0 \to \mu^+ \mu^-$ is given in
Fig.~4 as a function of $m_t$,$^{\Buras,\CampOdon}$ for $f_{B_s}=200$ MeV,
$\tau_{B_s} = 1.49$ psec, and $\vert V_{ts} \vert = 0.042$. For $m_t=150$
GeV, this gives
\begin{eqnarray}
BR (B_s^0\to \mu^+\mu^-) & = & 2 \times 10^{-9}~, \nonumber \\
BR (B_s^0\to \tau^+\tau^-) & = & 4 \times 10^{-7}~.
\end{eqnarray}

\vspace{7.7cm}
\centerline{Figure 4: Standard model branching ratio for $B_s^0 \to \mu^+
\mu^-$ as a function of $m_t$, }
\centerline{~~ assuming $f_{B_s}=200$ MeV, $\tau_{B_s}=1.49$ psec, and
$\vert V_{ts} \vert = 0.042$.}

In two-Higgs-doublet models, there can be an enhancement to the rate for
$B_s^0 \to \mu^+ \mu^-, \tau^+\tau^-$ by as much as one to two orders of
magnitude.$^\HiggsBmumu$ Since this decay proceeds through the loop-induced
exchange of a neutral Higgs scalar, the constraint on the $M_{{\sss
H}^\pm}$ from $\bsg$ is unimportant. In extended technicolour models
without a GIM mechanism, the rate can also be an order of magnitude bigger
than that of the SM.$^\Randall$ (Recall that extended technicolour models
with a GIM mechanism are already in conflict with date from $B\to
\mu^+\mu^-X$.) Finally, light leptoquarks could also enhance the rate for
$B_s^0 \to \mu^+ \mu^-$.

\newpage
\ttwohead{3.5}{$B_s\to\gamma\gamma$}

All that I will say about this process$^{\CampOdon,\aligreub}$ is that in
the SM $BR(B_s\to\gamma\gamma) = 1.5 \times 10^{-8}$ for $m_t=150$ GeV and
$f_{B_s}=200$ MeV.

\ttwohead{3.6}{Hadronic penguins}

The predictions for the exclusive rates of penguin-induced hadronic $B$
decays are highly model dependent. However, it is important to measure the
branching ratios of such decays for several reasons. First, this will give
us some idea as to the importance of penguin contributions$^\penguins$ in
CP-violating hadronic $B$ asymmetries. Also, we will be able to test
different models of exclusive decays and hence gain some information
regarding QCD effects in $B$ decays.

Some examples of such penguin-induced decays$^{\hadmodels,\Deshpande}$ and
their predicted branching ratios (taken from Ref.~\Deshpande) are given in
Table 1. These specific final states have been chosen since the signal
consists only of charged particles, so that these processes might be
observable at hadron colliders.

\begin{center}
\begin{tabular}{|c|c|}
\hline
Mode & Branching Ratio \\
\hline
$B^+ \to K^0 \pi^+$ & $1.06 \times 10^{-5}$ \\
\phantom{$B^+\to$} \negthinspace \negthinspace $K^+ \phi$ & $1.12 \times
10^{-5}$ \\
\phantom{$B^+\to$} $\; K^{*0} \pi^+$ & $0.58 \times 10^{-5}$ \\
\phantom{$B^+\to$} $K^{*+} \phi$ & $3.12 \times 10^{-5}$ \\
\hline
$B_d^0 \to K^{*0} \phi$ & $3.12 \times 10^{-5}$ \\
\phantom{$B_d^0 \to$} $\; K^{*0} \rho^0$ & $0.62 \times 10^{-5}$ \\
\hline
\end{tabular}
\end{center}
\centerline{Table 1: Some exclusive penguin-induced hadronic $B$ decays }
\centerline{\qquad\qquad and their predicted branching ratios (from
Ref.~\Deshpande).}

\onehead{4.}{$B_c$ PHYSICS}

One particularly interesting piece of $B$ physics which is likely to be
studied at hadron colliders is the $B_c = ({\bar b}c)$ system (for further
discussion regarding $B_c$ physics, see Refs.~\CHill\ and \CQuigg). The
mass of the $B_c$ has been calculated,$^{\CQuigg,\Bcmass}$ using potential
models, to be $\simeq 6.25$ GeV. Its production cross-section is about
$\sigma(B_c)/\sigma(b{\bar b}) \sim 10^{-3}$. This leads to$^\Bcproduction$
\begin{eqnarray}
1.3 \times 10^3 & & {\hbox{$B_c$'s per year ($10^7$ sec.) at LEP,}}
\nonumber\\
2.0 \times 10^6 & & {\hbox{\phantom{$B_c$'s per year ($10^7$ sec.) at}
TeVatron,}} \nonumber\\
1.1 \times 10^7 & & {\hbox{\phantom{$B_c$'s per year ($10^7$ sec.) at}
LHC (fixed target),}} \nonumber\\
1.1 \times 10^{11} & & {\hbox{\phantom{$B_c$'s per year ($10^7$ sec.) at}
LHC.}} \nonumber
\end{eqnarray}

\vspace{3.4cm}
\centerline{Figure 5: The three mechanisms for $B_c$ decay:}
\centerline{\qquad\qquad\qquad\qquad\qquad (i) $c$-spectator, (ii)
$b$-spectator, (iii) annihilation.}

\newpage

The main reason that $B_c$ mesons are so interesting is that there are
three mechanisms for their decay, shown in Fig.~5. Examples of these
different decays are:
\begin{eqnarray}
{\hbox{$c$-spectator:}} & & B_c^+ \to \Psi e^+ \nu_e ~, \nonumber\\
& & B_c^+ \to \eta_c e^+ \nu_e ~, \nonumber\\
& & B_c^+ \to \Psi \pi^+ ~,\nonumber\\
& & B_c^+ \to D^+ {\overline{D^0}} ~, \\
{\hbox{$b$-spectator:}} & & B_c^+ \to B_s^{(*)} e^+ \nu_e ~, \nonumber\\
& & B_c^+ \to B^{(*)} e^+ \nu_e ~, \nonumber\\
& & B_c^+ \to B_s^{(*)} \rho^+ ~,\nonumber\\
& & B_c^+ \to B^{(*)+} {\overline{K^0}} ~, \\
{\hbox{annihilation:}} & & B_c^+ \to \tau^+ \nu_\tau ~, \nonumber\\
& & B_c^+ \to D^{(*)+} K^0 ~.
\end{eqnarray}
Note that, unlike $B_u$'s, $B_d$'s and $B_s$'s, the annihilation decays of
the $B_c$ are expected to be important. There are a number of reasons for
this. First, helicity suppression is ineffective if there are heavy
particles (e.g.\ $\tau$, $D$, ...) in the final state. Second, in the $B_c$
system, such decays are unsuppressed by CKM factors. And finally, $f_{B_c}$
is expected to be large.

The relative importance of these three different decay mechanisms have been
estimated. Using quark and spectator models, and taking $\tau_{B_c}\simeq 5
\times 10^{-13}$ sec., the inclusive branching ratios for each of these
three types of decay are predicted to be:$^\Bcdecay$
\begin{eqnarray}
{\hbox{$c$-spectator:}} & & 37\%, \nonumber \\
{\hbox{$b$-spectator:}} & & 45\%, \nonumber \\
{\hbox{annihilation:}} & & 18\%.
\end{eqnarray}
Assuming $\tau_{B_c}\simeq 9 \times 10^{-13}$ sec., QCD sum rules
give:$^\BcdecayQCD$
\begin{eqnarray}
{\hbox{$c$-spectator:}} & & 48\%, \nonumber \\
{\hbox{$b$-spectator:}} & & 39\%, \nonumber \\
{\hbox{annihilation:}} & & 13\%.
\end{eqnarray}
For a more complete discussion of these relative inclusive branching
ratios, see Ref.~\CQuigg.

There are several particularly interesting decay modes of the $B_c$ which
involve a $\Psi$ in the final state. The decay $B_c^+ \to \Psi \pi^+$ is
likely to be the discovery mode. Its branching ratio is estimated to be $2
\times 10^{-3}$ and it permits the full reconstruction of the $B_c$. The
decay $B_c^+ \to \Psi \mu^+ \nu_\mu$ has a large branching ratio
(${\hbox{1--4}} \times 10^{-2}$) and its signal is three leptons coming from
the same vertex. In fact, $BR(B_c\to \Psi + X)$ is estimated to be
(19-24)\%, which means that the $B_c$ probably can be seen at CDF.

Given a sufficiently large sample of $B_c$'s, it is even possible to look
for CP violation in the $B_c$ system.$^\BcCP$ In order to have a non-zero
CP-violating decay-rate asymmetry, it is necessary to choose a final state
which can be reached via two different weak amplitudes. For example, the
decay $B_c^+ \to D^0 K^+$ has two contributions with different CKM matrix
elements -- a $c$-spectator tree diagram and a ${\bar b}\to {\bar s}$
penguin diagram. Another example is the processes $B_c^\pm \to D^0 D_s^\pm$
and $B_c^\pm \to {\overline{D^0}} D_s^\pm$. By measuring these decay rates
and the rate for $B_c^\pm \to D^0_{\sss CP} D_s^\pm$, where $D^0_{\sss CP}$
is identified by its decay to a CP eigenstate, the angle $\gamma$ of the
unitarity triangle can in principle be extracted.$^\GroWyler$
(Unfortunately, this particular example is probably experimentally
unfeasible, due to the tiny product branching ratios.)

\onehead{5.}{$\bs$-$\bsb$ MIXING}

The measurement of $\bs$-$\bsb$ mixing$^\xsreview$ is important for a
number of reasons:
\begin{itemize}
\item The mixing parameter $\xs\equiv (\Delta M)_{B_s}/\Gamma_{B_s}$ is
expected to be large ($>3$). If found to be small, this would be a smoking
gun for new physics.
\item $\xs$ can be used in conjuction with $x_d$ to
get a handle on $V_{td}$:
\beq
{x_d \over x_s} \sim {\fbb \over \fbbs} \left\vert {V_{td}\over V_{ts}}
\right\vert^2.
\label{xsxd}
\eeq
The ratio of hadronic matrix elements is usually known better than each
individual one. Thus, the measurement of $\xs$ would enable us to extract
$\vert V_{td} \vert$ with better precision.
\item An accurate knowledge of $\xs$ is needed to extract the CP-violating
angle $\gamma$ in $\bs$ decays.
\end{itemize}

In the SM, $\bs$-$\bsb$ mixing is dominated by $t$-quark exchange in the
box diagram, leading to
\beq
\xs = \tau_{B_s} \frac{G_F^2}{6\pi^2}M_W^2M_{B_s}\left(\fbbs\right)
\eta_{B_s} y_t f_2(y_t) \vert V_{ts}^*V_{tb}\vert^2~,
\eeq
in which $y_t \equiv m_t^2/M_{\sss W}^2$ and
\beq
f_2(x) = \frac{1}{4} + \frac{9}{4}\frac{1}{(1-x)}
- \frac{3}{2}\frac{1}{(1-x)^2}
- \frac{3}{2} \frac{x^2\ln x}{(1-x)^3}~.
\eeq
Taking
\begin{eqnarray}
\left\vert V_{ts} \right\vert = \left\vert V_{cb} \right\vert &=& 0.042 \pm
0.005~, \nonumber\\
\tau_{B_s} = \tau_B &=& 1.49 \pm 0.04~{\rm psec}~, \nonumber\\
\eta_{B_s} = \eta_B &=& 0.55 ~, \nonumber\\
M_{B_s} &=& 5.38~{\rm GeV}~,
\end{eqnarray}
this gives
\beq
\xs = \left(175\pm 21\right)\frac{\fbbs}{(1~{\rm GeV})^2} \, y_t f_2(y_t).
\eeq
For $89~{\rm GeV} \le m_t \le 182~{\rm GeV}$, the function $y_t f_2(y_t)$
is in the range 0.88--2.72, and is equal to 2.03 for the ``central'' value
of $m_t$, 150 GeV.

A consensus has not yet been reached regarding the value of $\fbbs$.
Potential models and QCD sum rules tend to give smaller values, while
lattice calculations give larger values. I will therefore consider two
ranges for $\fbbs$:
\begin{eqnarray}
(I):&~&~~\fbs\sqrt{B_{B_s}} = 180 \pm 35~{\rm MeV}, \nonumber \\
(II):&~&~~\fbs\sqrt{B_{B_s}} = 225 \pm 25~{\rm MeV}.
\end{eqnarray}
These lead to the following ``central'' values for $\xs$ (taking $m_t=150$
GeV):
\begin{eqnarray}
(I):~~\xs &=& 11.5, \nonumber \\
(II):~~\xs &=& 18.0.
\end{eqnarray}
The ``$1\sigma$'' lower limits on $\xs$ are
\begin{eqnarray}
(I):&~&~~\xs > 3.3, \nonumber \\
(II):&~&~~\xs > 6.6.
\end{eqnarray}

Clearly there is a large theoretical uncertainty regarding the hadronic
matrix elements. For example, lattice estimates give$^\fBlattice$
\begin{eqnarray}
f_{B_d} & = & {\hbox{188--246 MeV,}} \nonumber\\
f_{B_s} & = & {\hbox{204--241 MeV.}}
\end{eqnarray}
However, the error on the ratio of these two quantities is considerably
smaller:$^\fBratio$
\beq
{\fbbs \over \fbb} = 1.19 \pm 0.10~.
\eeq
This is why a precise measurement of $\xs$ can be used, along with $x_d$,
to extract $V_{td}$ (see Eq.~\ref{xsxd}).

It is possible to get smaller values of $\xs$ if one invokes physics beyond
the SM. Examples of such new physics are: a fourth generation,$^\fourthgen$
non-minimal SUSY models,$^\SUSYxs$ fine-tuned left-right symmetric
models$^\lrxs$ and models with $Z$-mediated flavour-changing neutral
currents.$^\zfcnc$ However, none of these is particularly compelling.

\onehead{6.}{T VIOLATION}

The last topic I wish to briefly discuss is T violation. By this I do not
mean CP violation, which is discussed elsewhere,$^\BCPviolation$ but rather
triple-product correlations. There are two examples of these which have
been discussed in the literature, having to do with the decays $B\to V_1
V_2$ ($V_1$ and $V_2$ are spin-1 mesons) and $B \to D^* \ell \nu_\ell$. I
won't go into very much detail regarding either of these decays, preferring
instead to simply sketch out the salient features.

Consider first the decay$^\Valencia$
\beq
B(p) \to V_1(k,\epsilon_1) V_2(q,\epsilon_2),
\eeq
in which the particles are specified by their 4-momenta ($p,k,q$) and their
polarizations ($\epsilon_1,\epsilon_2$). The most general decay amplitude
can be written
\beq
{\cal M} = a \, \epsilon_1\cdot\epsilon_2 + {b\over m_1 m_2} (p\cdot
\epsilon_1) (p\cdot \epsilon_2) + i {c\over m_1 m_2}
\epsilon^{\alpha\beta\gamma\delta} \epsilon_{1\alpha} \, \epsilon_{2\beta}
\, k_\gamma \, p_\delta~,
\eeq
in which
\begin{eqnarray}
a & = & \vert a \vert e^{i(\delta_a + \phi_a)}~, \nonumber\\
b & = & \vert b \vert e^{i(\delta_b + \phi_b)}~, \nonumber\\
c & = & \vert c \vert e^{i(\delta_c + \phi_c)}~,
\label{abc}
\end{eqnarray}
where $\delta_{a,b,c}$ and $\phi_{a,b,c}$ are the strong phases and the
weak phases, respectively. The corresponding amplitude for the decay of the
antiparticle is
\beq
{\overline{\cal M}} = {\bar a} \, \epsilon_1\cdot\epsilon_2 + {{\bar
b}\over  m_1 m_2} (p\cdot \epsilon_1) (p\cdot \epsilon_2) - i {{\bar
c}\over m_1 m_2} \epsilon^{\alpha\beta\gamma\delta} \epsilon_{1\alpha} \,
\epsilon_{2\beta} \, k_\gamma \, p_\delta~,
\eeq
in which ${\bar a}, {\bar b}, {\bar c}$ are identical to $a,b,c$
(Eq.~\ref{abc}), except that the $\phi_{a,b,c}$ change sign.

Now, the asymmetry
\beq
A_{\sss B} = {N_{\rm events} ({\vec k}\cdot {\vec\epsilon}_1 \times
{\vec\epsilon}_2 > 0 ) -
N_{\rm events} ({\vec k}\cdot {\vec\epsilon}_1 \times
{\vec\epsilon}_2 < 0 ) \over N_{\sss TOT} }
\eeq
can be written
\beq
A_{\sss B} \sim {\rm Im}(ac^*) \sim \vert a c \vert \sin (\delta+\phi),
\eeq
where $\delta\equiv\delta_a-\delta_c$ and $\phi\equiv\phi_a-\phi_c$. If we
imagine measuring a similar asymmetry $A_{\overline{\sss B}}$ for the
antiparticle decay, then we can obtain
\begin{eqnarray}
A_{\sss B} + A_{\overline{\sss B}} & \sim & \vert a c \vert \cos \delta
\sin \phi~, \nonumber\\
A_{\sss B} - A_{\overline{\sss B}} & \sim & \vert a c \vert \sin \delta
\cos \phi~.
\end{eqnarray}

The useful thing about such asymmetries, particularly the sum $A_{\sss B} +
A_{\overline{\sss B}}$, is that they are sensitive to the weak phases only,
i.e.\ they do not vanish if $\delta=0$. On the other hand, the question of
how to relate phases at the meson level to phases at the quark level, and
of how to calculate strong phases, introduces much theoretical uncertainty
and model dependence.$^\BBVmodel$ Still, the signals would be interesting
to look for. Some possible decay modes are: $\bdb\to\rho^{*+} K^{*-}$,
$B^-\to\Psi K^{*-}$ and $\bsb \to \Psi\phi$.

Another interesting process is the decay $B\to D^* \ell \nu_\ell$, in which
the $D^*$ decays further to $D\pi$.$^\Btriple$ The triple product ${\vec
p}_\ell \cdot ({\vec p}_{\sss D^*} \times {\vec p}_{\sss D})$ is
T-violating. There are a variety of asymmetries one can measure which
depend on this triple product (I refer the reader to Ref.~\Btriple\ for
more details). Again, to go from the quark-level calculation to the
meson-level measurement introduces hadronic uncertainties and model
dependence. However, this triple product vanishes in the SM, so that this
would be another way of looking for CP violation from new physics.


\begin{thebibliography}{7.}{99}

\bibitem{semilepmodels} G. Altarelli, M. Cabibbo, G. Corbo, L. Maiani and
G. Martinelli, \npb{208}{82}{365};
M. Wirbel, B. Stech and M. Bauer, \zpc{29}{85}{637};
J.G. K\"orner and G.A. Schuler, \zpc{38}{88}{511};
N. Isgur, D. Scora, B. Grinstein and M. Wise, \prd{39}{89}{799}.

\bibitem{HQET} N. Isgur and M.B. Wise, \plb{232}{89}{113}, {\bf B237}
(1990) 527.

\bibitem{hadmodels} M. Bauer, B. Stech and M. Wirbel, \zpc{34}{87}{103};

\bibitem{gronwak} M. Gronau and S. Wakaizumi, \prl{68}{92}{1814}.

\bibitem{otherproc}
For a review of HQET, see B. Grinstein, these proceedings; \\
for a discussion of $B$ baryons, see B. Kayser, these proceedings.

\bibitem{langacker} P. Langacker and S.U. Sankar, \prd{40}{89}{1569}.

\bibitem{houwyler} W.-S. Hou and D. Wyler, \plb{292}{92}{364}.

\bibitem{Afb} J.G. K\"orner and G.A. Schuler, \plb{226}{89}{185};
F.J. Gilman and R.L. Singleton, Jr., \prd{41}{90}{142};
M. Gronau and S. Wakaizumi, \plb{280}{92}{79};
S. Sanghera et.\ al.\ (CLEO Collaboration), \prd{47}{93}{791}.

\bibitem{LEPrightB} J.F. Amundson, J.L. Rosner, M. Worah and M.B. Wise,
\prd{47}{93}{1260}; Z. Hioki, \plb{303}{93}{125}.

\bibitem{bsgQCD}
S. Bertolini, F. Borzumati and A. Masiero, \prl{59}{87}{180};
N.G. Deshpande et.\ al., \prl{59}{87}{183};
B. Grinstein, R. Springer and M.B. Wise, \plb{202}{88}{138},
\npb{339}{90}{269};
R. Grigjanis et.\ al., \plb{213}{88}{355}, {\bf 224} (1989) 209, {\bf 286}
(1992) 413(E);
G. Cella et.\ al., \plb{248}{90}{181};
A. Ali and C. Greub, \zpc{49}{91}{431}, \plb{287}{92}{191};
P. Cho and B. Grinstein, \npb{365}{91}{279};
M. Misiak, \npb{393}{93}{23}.

\bibitem{ali} A. Ali, in {\em Proceedings of the 1991 ICTP, Trieste, Summer
School in High Energy Physics and Cosmology} (World Scientific, Singapore,
1992), p.~153.

\bibitem{CLEO}
1991: M. Battle et.\ al.\ (CLEO Collaboration), in {\em Proceedings of the
Joint International Lepton-Photon Symposium \& Europhysics Conference on
High Energy Physics}, eds. S. Hegarty, K. Potter and E. Quercigh (World
Scientific, Singapore, 1992), p.~869; \\
1993: E. Thorndike (CLEO Collaboration), talk given at the {\em 1993
Meeting of the American Physical Society}, Washington, D.C., April, 1993;
R. Ammar et.\ al.\ (CLEO Collaboration), \prl{71}{93}{674}.

\bibitem{Hewett} J.L. Hewett, \prl{70}{93}{1045}.

\bibitem{Barger} V. Barger, M.S. Berger and R.J.N. Phillips,
\prl{70}{93}{1368}.

\bibitem{BelangerGeng} G. B\'elanger, C.Q. Geng and P. Turcotte,
UdeM-LPN-TH-93-148 (1993).

\bibitem{SUSYcorrs} M.A. D{\'\i}az, \prd{48}{93}{2152}, \plb{304}{93}{278}.

\bibitem{SUSYbsg} N. Oshimo, \npb{404}{93}{20};
T. Hayashi, M. Matsuda and M. Tanimoto, KU-01-93, AUE-01-93, EHU-01-93
(1993);
R. Barbieri and G.F. Giudice, \plb{309}{93}{86};
J.L. Lopez, D. Nanopoulos and G.T. Park, \prd{48}{93}{974};
Y. Okada, \plb{315}{93}{119};
R. Garisto and J.N. Ng, TRI-PP-93-66 (1993);
F. Borzumati, DESY 93-090 (1993).

\bibitem{ARGUSCLEO} Particle Data Group, {\em Phys. Rev.} {\bf D45} (1992)
Vol.~45, part  II.

\bibitem{bsllnunurate}
W.-S. Hou, R.S. Willey and A. Soni, \prl{58}{87}{1608}.

\bibitem{bsllrate}
N.G. Deshpande and J. Trampetic, \prl{60}{88}{2583};
C.S. Lim, T. Morozumi and A.I. Sanda, \plb{218}{89}{343};
B. Grinstein, M.J. Savage and M.B. Wise, \npb{319}{89}{271};
N.G. Deshpande, J. Trampetic and K. Panose, \prd{39}{89}{1461};
W. Jaus and D. Wyler, \prd{41}{90}{3405}.

\bibitem{aligreub} A. Ali, C. Greub and T. Mannel, in {\em ECFA Workshop on
a European $B$-Meson Factory}, eds.\ R. Aleksan and A. Ali (1993), p.~155.

\bibitem{UA1} C. Albajar et.\ al., UA1 Collaboration, \plb{262}{91}{163}.

\bibitem{AliMor} A. Ali, T. Mannel and T. Morozumi, \plb{273}{91}{505}.

\bibitem{BBM} S. Bertolini, F. Borzumati and A. Masiero, in {\em $B$
Decays}, ed. S. Stone (World Scientific, Singapore, 1992), p.~458, and
references therein.

\bibitem{Fran} F. Borzumati, private communication.

\bibitem{GrinNir} B. Grinstein, Y. Nir and J.M. Soares, SSCL-Preprint-482,
WIS-93/67/July-PH, CMU-HEP93-10; DOE-ER/40682-35 (1993).

\bibitem{Buras} A.J. Buras and G. Buchalla, \npb{400}{93}{225}.

\bibitem{CampOdon} B.A. Campbell and P.J. O'Donnell, \prd{25}{82}{1989}.

\bibitem{HiggsBmumu} M.J. Savage, \plb{266}{91}{135}; W. Skiba and J.
Kalinowski, \npb{404}{93}{3}.

\bibitem{Randall} L. Randall and R. Sundrum, \plb{312}{93}{148}.

\bibitem{penguins} D. London and R.D. Peccei, \plb{223}{89}{257};
M. Gronau, \prl{63}{89}{1451}; B. Grinstein, \plb{229}{89}{280}.

\bibitem{Deshpande} N.G. Deshpande and J. Trampetic, \prd{41}{90}{895}.

\bibitem{CHill} C. Hill, these proceedings

\bibitem{CQuigg} C. Quigg, these proceedings.

\bibitem{Bcmass} A.K. Likhoded, S.R. Slabospitsky, M. Mangano, and G.
Nardulli, BARI-TH/93-137 and references therein.

\bibitem{Bcproduction}
M. Lusignoli, M. Masetti and S. Petrarca, \plb{266}{91}{142};
C.-H. Chang, Y.-Q. Chen, \plb{284}{92}{127}, \prd{46}{92}{3845};
E. Braaten, K. Cheung and T.C. Yuan, NUHEP-TH-93-6, UCD-93-9 (1993).

\bibitem{Bcdecay} V.V. Kiselev and A.V. Tkabladze,
{\em Sov.~J.~Nucl.~Phys.} {\bf 48} (1988) 536;
M. Lusignoli and M. Masetti, \zpc{51}{91}{549}.

\bibitem{BcdecayQCD} P. Colangelo, G. Nardulli and N. Paver,
\zpc{57}{93}{43}.

\bibitem{BcCP} M. Masetti, \plb{286}{92}{160}.

\bibitem{GroWyler} M. Gronau and D. Wyler, \plb{253}{91}{483}.

\bibitem{xsreview} For a review of $\bs$-$\bsb$ mixing, see A. Ali and D.
London, {\em J. Phys.\ G: Nucl.\ Part.\ Phys.} {\bf 19} (1993) 1069.

\bibitem{fBlattice} C. Alexandrou et.\ al., \npb{374}{92}{263}.

\bibitem{fBratio} A. Abada et.\ al., \npb{376}{92}{172}.

\bibitem{fourthgen} D. London, \plb{234}{90}{354}.

\bibitem{SUSYxs} I.I. Bigi and F. Gabbiani, \npb{352}{91}{309}, and
references therein.

\bibitem{lrxs} D. London and D. Wyler, \plb{232}{89}{503}.

\bibitem{zfcnc} Y. Nir and D. Silverman, \prd{42}{90}{1477};
D. Silverman, \prd{45}{92}{1800}.

\bibitem{BCPviolation} For a review of CP violation in the $B$ system, see
M. Gronau, these proceedings.

\bibitem{Valencia} G. Valencia, \prd{39}{89}{3339}.

\bibitem{BBVmodel} For an example of such a model analysis, see G. Kramer
and W.F. Palmer, \prd{45}{92}{193}, \plb{279}{92}{181}, \prd{46}{92}{2969},
DESY 92-043 (1992); G. Kramer, T. Mannel and W.F. Palmer,
\zpc{55}{92}{497}.

\bibitem{Btriple} E. Golowich and G. Valencia, \prd{40}{89}{112};
J.G. K\"orner, K. Schilcher and Y.L. Wu, \plb{242}{90}{119}.

\end{thebibliography}
\end{document}